\documentclass{article}
\usepackage{amsfonts, amsmath, amssymb}
\usepackage[T1]{fontenc} %
\usepackage[latin1]{inputenc} %
\usepackage[dvips]{graphicx}
\usepackage{color}
\newtheorem{lemma}{Lemma}
\newtheorem{theorem}{Theorem}
\newtheorem{coro}{Corollary}

\newtheorem{obs}{Observation}

\def\RMO{R{\footnotesize{OOTED}} M{\footnotesize{AXIMUM}} L{\footnotesize{EAF}} O{\footnotesize{UTBRANCHING}} }
\def\RMOB{R{\footnotesize{OOTED}} M{\footnotesize{AXIMUM}} L{\footnotesize{EAF}} O{\footnotesize{UTBRANCHING}}. }
\def\RMOC{R{\footnotesize{OOTED}} M{\footnotesize{AXIMUM}} L{\footnotesize{EAF}} O{\footnotesize{UTBRANCHING}}: }
\def\RMOD{R{\footnotesize{OOTED}} M{\footnotesize{AXIMUM}} L{\footnotesize{EAF}} O{\footnotesize{UTBRANCHING}}, }
\def\MO{M{\footnotesize{AXIMUM}} L{\footnotesize{EAF}} O{\footnotesize{UTBRANCHING}}}
\def\ML{M{\footnotesize{AXIMUM}} L{\footnotesize{EAF}} S{\footnotesize{PANNING}} T{\footnotesize{REE}}}

\title{On finding directed trees with many leaves}

\author{Jean Daligault and St\'ephan Thomass\'e}
\begin{document}
\maketitle

\begin{abstract}
The \RMO problem consists in finding a spanning directed tree rooted at some prescribed vertex of a 
digraph with the maximum number of leaves. Its parameterized version asks if there 
exists such a tree with at least $k$ leaves. We use the notion of $s-t$ numbering studied in \cite{stnum}, \cite{stnumdir}, \cite{LLWemb} to exhibit combinatorial bounds on the existence of spanning directed trees with many leaves. These combinatorial bounds allow us to produce a constant factor approximation algorithm for finding directed trees with many leaves, whereas the best known approximation algorithm has a $\sqrt{OPT}$-factor \cite{DrescherMaxLeaf}. We also show that \RMO admits a quadratic kernel, improving over the cubic kernel given by Fernau et al \cite{FernauMaxLeaf}.
\end{abstract}

\section{Introduction}
An \emph{outbranching} of a digraph $D$ is a spanning directed tree in $D$
. 
We consider the following problem:

\vspace{12pt}
{\bf \RMOC}

\vspace{12pt}
\textbf{Input}: A digraph $D$, an integer $k$, a vertex $r$ of $D$.

\textbf{Output}: TRUE if there is an outbranching of $D$ rooted at $r$ with at least $k$ leaves, otherwise FALSE.
\vspace{12pt}

This problem is equivalent to finding a Connected Dominating Set of size at most $|V(D)|-k$, connected meaning in this setting that every vertex is reachable by a directed path from $r$. Indeed, the set of internal nodes in an outbranching correspond to a connected dominating set.

Finding undirected trees with many leaves has many applications in the area of communication networks, see \cite{MaxLeafApplication2} or \cite{MaxLeafApplication} for instance. An extensive litterature is devoted to the paradigm of using a small connected dominating set as a backbone for a communication network.

\RMO is NP-complete, even restricted to acyclic digraphs \cite{AlonMaxLeaf2}, and MaxSNP-hard, even on undirected graphs \cite{ApxHardMaxLeaf}.

Two natural ways to tackle such a problem are, on the one hand, polynomial-time approximation algorithms, and on the other hand, parameterized complexity. Let us give a brief introduction on the parameterized approach.

\vspace{12pt}

An efficient way of dealing with NP-hard problems
is to identify a parameter which contains its 
computational hardness. For instance, instead 
of asking for a minimum vertex cover in a graph -
a classical NP-hard optimization question - one
can ask for an algorithm which would decide, in
$O(f(k).n^d)$ time for some fixed $d$, if a graph 
of size $n$ has a  vertex cover of size at 
most $k$. If such an algorithm exists, the problem
is called {\it fixed-parameter tractable}, or FPT for 
short. An extensive literature is devoted to FPT, 
the reader is invited to read \cite{DowneyFellows}, \cite{FlumGrohe}
and \cite{NiedBookFPT}.

Kernelization is a natural way of proving that a 
problem is FPT. Formally, a {\it kernelization algorithm}
receives as input an instance $(I,k)$ of the parameterized
problem, and outputs, in polynomial time  in the size of 
the instance, another instance $(I',k')$ such that: $k'\leq k$, the size of $I'$ only depends of $k$, and the instances $(I,k)$ and $(I',k')$ are both true or both false.


The reduced instance $(I',k')$ is called a {\it kernel}.
The existence of a kernelization algorithm 
clearly implies the FPT character of the problem since 
one can kernelize the instance, and then solve the reduced
instance $G',k'$ using brute force, hence giving an 
$O(f(k)+n^d)$ algorithm. 
A classical result asserts that being FPT is indeed equivalent to 
having kernelization. The drawback of this result is that 
the size of the reduced instance $G'$ is not necessarily 
small with respect to $k$. A much more constrained condition
is to be able to reduce to an instance of polynomial 
size in terms of $k$. Consequently, in the zoology of parameterized 
problems, the first distinction is done between three classes: 
W[1]-hard, FPT, polykernel. 

A kernelization algorithm can be used as a preprocessing step to reduce the size of the instance before applying some other parameterized algorithm. Being able to ensure that this kernel has actually polynomial size in $k$ enhances the overall speed of the process. See \cite{GuoNiedFPT} for a recent review on kernelization.

\vspace{12pt}

An extensive litterature is devoted to finding trees with many leaves in undirected and directed graphs. The undirected version of this problem, \ML, has been extensively studied. There is a factor 2 approximation algorithm for the \ML ~problem \cite{SolisMaxLeafApprox}, and a $3.75k$ kernel \cite{EstivillMaxLeaf}. An $O^*(1,94^n)$ exact algorithm was designed in \cite{FominKratschMaxLeaf}. Other graph theoretical results on the existence of trees with many leaves can be found in \cite{SeymourMaxLeaf} and \cite{StorerMaxLeaf}.

The best approximation algorithm known for \MO ~is a factor $\sqrt{OPT}$ algorithm \cite{DrescherMaxLeaf}. From the Parameterized Complexity viewpoint, Alon et al showed that \MO ~restricted to a wide class of digraphs containing all strongly connected digraphs is FPT \cite{AlonMaxLeaf}, and Bonsma and Dorn extended this result to all digraphs and gave a faster parameterized algorithm \cite{BonsmaMaxLeaf}. Very recently, Kneis, Langer and Rossmanith \cite{KneisMaxLeaf} obtained an $O^*(4^k)$ algorithm for \MO, which is also an improvement for the undirected case over the numerous FPT algorithms designed for \ML. Fernau et al \cite{FernauMaxLeaf} proved that \RMO has a polynomial kernel, exhibiting a cubic kernel. They also showed that the unrooted version of this problem admits no polynomial kernel, unless polynomial hierarchy collapses to third level, using a breakthrough lower bound result by Bodlaender et al \cite{NoKernel}. A linear kernel for the acyclic subcase of \RMO and an $O^*(3,72^k)$ algorithm for \RMO were exhibited in \cite{GutinMaxLeaf}.

\vspace{12pt}
This paper is organized as follows. In Section~\ref{sbounds} we exhibit combinatorial bounds on the problem of finding an outbranching with many leaves. We use the notion of \emph{$s-t$ numbering} introduced in \cite{stnum}. We next present our reduction rules, which are independent of the parameter, and in the following section we prove that these rules give a quadratic kernel. We finally present a constant factor approximation algorithm in Section~\ref{approx} for finding directed trees with many leaves.

\section{Combinatorial Bounds}\label{sbounds}
Let $D$ be a directed graph. For an arc $(u,v)$ in $D$, we say that $u$ is an \emph{in-neighbour} of $v$, that $v$ is an \emph{outneighbour} of $u$, that $(u,v)$ is an \emph{in-arc} of $v$ and an \emph{out-arc of $u$}. The \emph{outdegree} of a vertex is the number of its outneighbours, and its \emph{indegree} is the number of its in-neighbours.
An outbranching with a maximum number of leaves is said to be \emph{optimal}. Let us denote by $maxleaf(D)$ the number of leaves in an optimal outbranching of $D$.

Without loss of generality, we restrict ourselves to the following. We exclusively consider loopless digraphs with a distinguished vertex of indegree 0, denoted by $r$. \emph{We assume that there is no arc $(u,r)$ in $D$ with $u\in V(D)$, and no arc $(x,y)$ with $x\neq r$ and $y$ an outneighbour of $r$, and that $r$ has outdegree at least 2}. Throughout this paper, we call such a digraph a \emph{rooted digraph}. Definitions will be made exclusively with respect to rooted digraphs, hence the notions we present, like connectivity and resulting concepts, do slightly differ from standard ones.
Let $D$ be a rooted digraph with a specified vertex $r$. 

The rooted digraph $D$ is \emph{connected} if every vertex of $D$ is reachable by a directed path rooted at $r$ in $D$. A \emph{cut} of $D$ is a set $S\subseteq V(D)-r$ such that there exists a vertex $z\notin S$ endpoint of no directed path from $r$ in $D-S$. We say that $D$ is \emph{2-connected} if $D$ has no cut of size at most 1. A cut of size 1 is called a \emph{cutvertex}. Equivalently, a rooted digraph is 2-connected if there are two internally vertex-disjoint paths from $r$ to any vertex besides $r$ and its outneighbours.

We will show that the notion of $s-t$ numbering behaves well with respect to outbranchings with many leaves. It has been introduced in \cite{stnum} for 2-connected undirected graphs, and generalized in \cite{stnumdir} by Cheriyan and Reif for digraphs which are 2-connected in the usual sense. We adapt it in the context of rooted digraphs.

Let $D$ be a 2-connected rooted digraph. An $r-r$ numbering of $D$ is a linear ordering $\sigma$ of $V(D)-r$ such that, for every vertex $x\neq r$, either $x$ is an outneighbour of $r$ 
or there exist two in-neighbours $u$ and $v$ of $x$ such that $\sigma(u)<\sigma(x)<\sigma(v)$. An equivalent presentation of an $r-r$ numbering of $D$ is an injective embedding $f$ of the graph $D$ where $r$ has been duplicated into two vertices $r_1$ and $r_2$, into the $[0,1]$-segment of the real line, such that $f(r_1)=0$, $f(r_2)=1$, and such that the image by $f$ of every vertex besides $r_1$ and $r_2$ lies inside the convex hull of the images of its in-neighbours. Such \emph{convex embeddings} have been defined and studied in general dimension by Lov\'asz, Linial and Wigderson in \cite{LLWemb} for undirected graphs, and in \cite{stnumdir} for directed graphs.

Given a linear order $\sigma$ on a finite set $V$, we denote by $\bar{\sigma}$ the linear order on $V$ which is the reverse of $\sigma$. An arc $uv$ of $D$ is a \emph{forward} arc
if $u=r$ or if $u$ appears before $v$ in $\sigma$; $uv$ is a \emph{backward} arc if $u=r$ or if $u$ appears after $v$ in $\sigma$. A spanning out-tree $T$ is \emph{forward} if all its arcs are forward. Similar
definition for \emph{backward} out-tree.

The following result and proof is just an adapted version of \cite{stnumdir},  given here for the sake of completeness.
 
\begin{lemma}\label{rrnum}
Let $D$ be a 2-connected rooted digraph. There exists an $r-r$ numbering of $D$.
\end{lemma}
\emph{Proof}: By induction over $D$. We first reduce to the case where the indegree of every vertex besides $r$ is exactly 2. Let $x$ be a vertex of indegree at least 3 in $D$. Let us show that there exists an in-neighbour $y$ of $x$ such that the rooted digraph $D-(y,x)$ is 2-connected.
Indeed, there exist two internally vertex disjoint paths from $r$ to $x$. Consider such two paths intersecting $N^-(x)$ only once each, and denote by $D'$ the rooted digraph obtained from $D$ by removing one arc $(y,x)$ not involved in these two paths. There are two internally disjoint paths from $r$ to $x$ in $D'$. Consider $z\in V(D)-r-x$. Assume by contradiction that there exists a vertex $t$ which cuts $z$ from $r$ in $D'$. As $t$ does not cut $z$ from $r$ in $D$ and the arc $(y,x)$ alone is missing in $D'$, $t$ must cut $x$ and not $y$ from $r$ in $D'$. Which is a contradiction, as there are two internally disjoint paths from $r$ to $x$ in $D'$. By induction, $D'$ has an $r-r$ numbering, which is also an $r-r$ numbering for $D$.

Hence, let $D$ be a rooted digraph, where every vertex besides $r$ has indegree $2$. As $r$ has indegree 0, there exists a vertex $v$ with outdegree at most 1 in $D$ by a counting argument. If $v$ has outdegree 0, then let $\sigma$ be an $r-r$ numbering of $D-v$, let $u_1$ and $u_2$ be the two in-neighbours of $v$. Insert $v$ between $u_1$ and $u_2$ in $\sigma$ to obtain an $r-r$ numbering of $D$.
Assume now that $v$ has a single outneighbour $u$. Let $w$ be the second in-neighbour of $u$. Let $D'$ be the graph obtained from $D$ by contracting the arc $(v,u)$ into a single vertex $uv$. As $D'$ is 2-connected, consider by induction an $r-r$ numbering $\sigma$ of $D'$. Replace $uv$ by $u$. It is now possible to insert $v$ between its two in-neighbours in order to make it so that $u$ lies between $v$ and $w$. Indeed, assume without loss of generality that $w$ is after $uv$ in $\sigma$. Consider the smallest in-neighbour $t$ of $v$ in $\sigma$. As $\sigma$ is an $r-r$ numbering of $D'$, $t$ lies before $uv$ in $\sigma$. We insert $v$ just after $t$ to obtain an $r-r$ numbering of $D$. $\square$
\\

Note that an $r-r$ numbering $\sigma$ of $D$ naturally gives two acyclic covering subdigraphs of $D$, the rooted digraph $D_{|\sigma}$ consisting of the forward arcs of $D$, and the rooted digraph $D_{|\bar{\sigma}}$ consisting of the backward arcs of $D$. The intersection of these two acyclic digraphs is the set of out-arcs of $r$.

\begin{coro}
Let $D$ be a 2-connected rooted digraph. There exists an acyclic connected spanning subdigraph $A$ of $D$ which contains at least half of the arcs of $D-r$.
\end{coro}
Let $G$ be an undirected graph. A \emph{vertex cover} of $G$ is a set of vertices covering all edges of $G$. A \emph{dominating set} of $G$ is a set $S\subseteq V$ such that for every vertex $x\notin S$, $x$ has a neighbour in $S$. A \emph{strongly dominating set} of $G$ is a set $S\subseteq V$ such that every vertex has a neighbour in $S$.

Let $D$ be a rooted digraph. A \emph{strongly dominating set} of $D$ is a set $S\subseteq V$ such that every vertex besides $r$ has an in-neighbour in $S$.
We need the following folklore result:
\begin{lemma}\label{domi1}
Any undirected graph $G$ on $n$ vertices and $m$ arcs has a vertex cover of size $\frac{n+m}{3}$.
\end{lemma}
\emph{Proof}: By induction on $n+m$. If there exists a vertex of degree at least 2 in $G$, choose it in the vertex cover, otherwise choose any non-isolated vertex.
$\square$

\begin{lemma}\label{domi2}
Let $G$ be a bipartite graph over $A\cup B$, with $d(a)= 2$ for every $a\in A$. There exists a subset of $B$ dominating $A$ with size at most $\frac{|A|+|B|}{3}$.
\end{lemma}
\emph{Proof}:
Let $G'$ be the graph which vertex set is $B$, and where $(b,b')$ is an arc if $b$ and $b'$ share a common neighbour in $A$. The result follows from Lemma~\ref{domi1} since $G'$ has $|A|$ arcs and $|B|$ vertices.
$\square$

\begin{coro}\label{leaves}
Let $D$ be an acyclic rooted digraph with $l$ vertices of indegree at least 2 and with a root of outdegree $d(r)\ge 2$. Then $D$ has an outbranching with at least 
$\frac{l+d(r)-1}{3}+1$ leaves.
\end{coro}
\emph{Proof}: Denote by $n$ the number of vertices of $D$. For every vertex $v$ of indegree at least 3, delete incoming arcs until $v$ has indegree exactly 2. Since $D$ is acyclic, it has a sink $s$. 

Let $Z$ be the set of vertices of indegree 1 in $D$, of size $n-1-l$. Let $Y$ be the set of in-neighbours of vertices of $Z$, of size at most $n-1-l$. Let $A'$ be the set of vertices of indegree 2 dominated by $Y$. Let $B=V(D)-Y-s$. Let $A$ be the set of vertices of indegree 2 not dominated by $Y$. Note that $Y$ cannot have the same size as $Z\cup A'$. Indeed, $Z$ contains the outneighbours of $r$, and hence $Y$ contains $r$, which has outdegree at least 2. More precisely, $|Y|+d(r)-1\le|Z\cup A'|$. As $B=V(D)-Y-s$ and $A=V(D)-A'-Z-r$, we have that $|B|\ge|A|+d(r)-1$. Moreover, as $Y$ has size at most $n-1-l$, we have that $|B|\ge l$. Consider a copy $A_1$ of $A$ and a copy $B_1$ of $B$. Let $G$ be the bipartite graph with vertex bipartition $(A_1,B_1)$, and where $(b,a)$, with $a\in A_1$ and $b\in B_1$, is an edge if $(b,a)$ is an arc in $D$. By Lemma~\ref{domi2} applied to $G$, there exists a set $X\subseteq B$ of size at most $\frac{|A|+|B|}{3}\le\frac{2|B|-(d(r)-1)}{3}$ which dominates $A$ in $D$. 
The set $C=X\cup Y$ strongly dominates $V(D)-r$ in $D$, and has size at most $|X|+|Y|\le \frac{2|B|-(d(r)-1)}{3}+|Y|= |B|+|Y|-\frac{|B|+d(r)-1}{3}$. As $|Y|+|B|=n-1$ and $|B|\ge l$, this yields $|X\cup Y| \le n-1-\frac{l+d(r)-1}{3}$. As $D$ is acyclic, any set strongly dominating $V-r$ contains $r$ and is a connected dominating set. Hence there exists an outbranching $T$ of $D$ having a subset of $C$ as internal vertices. $T$ has at least $\frac{l+d(r)-1}{3}+1$ leaves. 

$\square$\\

This bound is tight up to one leaf. 
The rooted digraph $D_k$ depicted in Figure 1 is 2-connected, has $3k-2$ vertices of indegree at least 2, $d(r)=3$ and $maxleaf(D_k)=k+2$.

\vspace{12pt}

\begin{figure}
\centering
\includegraphics[angle=-90]{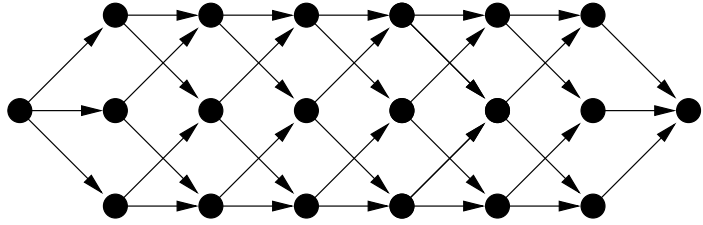}
\caption{The "boloney" graph $D_6$}
\end{figure}

Finally, the following combinatorial bound is obtained:

\begin{theorem}\label{bound1}
Let $D$ be a 2-connected rooted digraph with $l$ vertices of indegree at least 3. Then  $maxleaf(D)\ge\frac{l}{6}$.
\end{theorem}
\emph{Proof}: Apply Corollary~\ref{leaves} to the rooted digraph with the larger number of vertices of indegree 2 among $D_{\sigma}$ and $D_{\bar{\sigma}}$. $\square$\\

An arc is \emph{simple} if does not belong to a 2-circuit. A vertex $v$
is \emph{nice} if it is incident
to a simple in-arc. 

The second combinatorial bound is the following:
\begin{theorem}\label{bound2}
Let $D$ be 2-connected rooted digraph. Assume that $D$ has $l$ 
nice vertices. Then $D$ has an outbranching with at least $\frac{l}{24}$ leaves.
\end{theorem}
\emph{Proof}:
By Lemma~\ref{rrnum}, we consider an $r-r$ numbering $\sigma$ of $D$.
For every nice vertex $v$
(incident
to some in-arc $a$) with indegree at least three, delete incoming arcs
of $v$ different from $a$ until $v$ has only one incoming forward arc
and one incoming backward arc. For every other vertex of indegree at
least 3 in $D$, delete incoming arcs of $v$ until $v$ has only one
incoming forward arc and one incoming backward arc. At the end of this
process, $\sigma$ is still an $r-r$ numbering of the digraph $D$, and
the number of nice vertices has not decreased.

Denote by $T_f$ the set of forward arcs of $D$, and by $T_b$ the set of
backward arcs of $D$. As $\sigma$ is an $r-r$ numbering of $D$, $T_f$
and $T_b$ are spanning trees of $D$ which partition the arcs of $D-r$.

The crucial definition is the following: say that an arc $uv$ of $T_f$
(resp. of $T_b$), with $u\neq r$, is \emph{transverse} if $u$ and $v$
are \emph{incomparable} in $T_b$ (resp. in $T_f$), that is if $v$ is not
an ancestor of $u$ in $T_b$ (resp. in $T_f$). Observe that $u$ cannot be
an ancestor of $v$ in $T_b$ (resp. in $T_f$) since $T_b$ is backward (resp. $T_f$ is forward) while $uv$
is forward (resp. backward) and $u\neq r$.

Assume without loss of generality that $T_f$ contains more transverse
arcs than $T_b$. Consider now any planar
drawing of the rooted tree $T_b$. We will make use of this drawing to
define the following:
if two vertices $u$ and $v$ are incomparable in $T_b$, then one of these
vertices is to the left of the other, with respect to our drawing.
Hence, we can
partition the transverse arcs of $T_f$ into two subsets: the set $S_l$
of transverse arcs
$uv$ for which $v$ is to the left of $u$, and the set $S_r$ of
transverse arcs
$uv$ for which $v$ is to the right of $u$. Assume without loss of
generality that $|S_l|\ge |S_r|$.

The digraph $T_b\cup S_l$ is an acyclic digraph by definition of $S_l$.
Moreover,
it has $|S_l|$ vertices of indegree two since the heads of the arcs of
$|S_l|$
are pairwise distinct. Hence, by Corollary~\ref{leaves}, $T_b\cup S_l$
has an outbranching
with at least $\frac{|S_l|+d(r)-1}{3}+1$ leaves, hence so does $D$.

We now give a lower bound on the number of transverse arcs in $D$ to
bound $|S_l|$. Consider
a nice vertex $v$ in $D$, which is not an outneighbour of $r$, and with a simple in-arc $uv$ belonging to, say,
$T_f$. If $uv$ is not
a transverse arc, then $v$ is an ancestor of $u$ in $T_b$. Let $w$ be
the outneighbor of
$v$ on the path from $v$ to $u$ in $T_b$. Since $uv$ is simple, the
vertex $w$ is distinct
from $u$. No path in $T_f$ goes from $w$ to $v$, hence $vw$ is a
transverse arc. Therefore,
we proved that $v$ (and hence every nice vertex) is incident to a
transverse arc (either
an in-arc, or an out-arc). Thus there are at least $\frac{l-d(r)}{2}$
transverse arcs in $D$.

Finally, there are at least $\frac{l-d(r)}{4}$ transverse arcs in $T_f$, and thus
$|S_l|\ge \frac{l-d(r)}{8}$. In all,
$D$ has an outbranching with at least $\frac{l}{24}$ leaves.$\square$




\vspace{12pt}


As a corollary, the following result holds for oriented graphs (digraphs with no 2-circuit):
\begin{coro}
Every 2-connected rooted oriented graph on $n$ vertices has an outbranching with at least $\frac{n-1}{24}$ leaves.
\end{coro}

\section{Reduction Rules}\label{srules}
 We say that $P=\{x_1,\dots,x_l\}$, with $l\ge 3$, is a \emph{bipath of length $l-1$} if the set of arcs adjacent to $\{x_2,\dots,x_{l-1}\}$ in $D$ is exactly $\{(x_i,x_{i+1}),(x_{i+1},x_i) |i\in\{1,\dots,l-1\}\}$
.

To exhibit a quadratic kernel for \RMOD we use the following four reduction rules:
\begin{itemize}
\item[(0)] If there exists a vertex not reachable from $r$ in $D$, then reduce to a trivially FALSE instance. 
\item[(1)] Let $x$ be a cutvertex of $D$. Delete vertex $x$ and add an arc $(v,z)$ for every $v\in N^-(x)$ and $z\in N^+(x)-v$.
\item[(2)] Let $P$ be a bipath of length 4. Contract two consecutive internal vertices of $P$. 
\item[(3)] Let $x$ be a vertex of $D$. If there exists $y\in N^-(x)$ such that $N^-(x)-y$ cuts $y$ from $r$, then delete the arc $(y,x)$.
\end{itemize}

Note that these reduction rules are not parameter dependent. Rule (0) only needs to be applied once.

\begin{obs}\label{obs1}
Let $S$ be a cutset of a rooted digraph $D$. Let $T$ be an outbranching of $D$. There exists a vertex in $S$ which is not a leaf in $T$.
\end{obs}

\begin{lemma}\label{rules}
The above reduction rules are safe and can be checked and applied in polynomial time.
\end{lemma}
\emph{Proof}:
\begin{itemize}
\item[(0)] Reachability can be tested in linear time.
\item[(1)] Let $x$ be a cutvertex of $D$. Let $D'$ be the graph obtained from $D$ by deleting vertex $x$ and adding an arc $(v,z)$ for every $v\in N^-(x)$ and $z\in N^+(x)-v$. Let us show that maxleaf$(D)$ $=$ maxleaf$(D')$. Assume $T$ is an outbranching of $D$ rooted at $r$ with $k$ leaves. By Observation~\ref{obs1}, $x$ is not a leaf of $T$. Let $f(x)$ be the father of $x$ in $T$. Let $T'$ be the tree obtained from $T$ by contracting $x$ and $f(x)$. $T'$ is an outbranching of $D'$ rooted at $r$ with $k$ leaves. 

Let $T'$ be an outbranching of $D'$ rooted at $r$ with $k$ leaves. $N^-(x)$ is a cut in $D'$, hence by Observation~\ref{obs1} there is a non-empty collection of vertices $y_1,\dots,y_l\in N^-(x)$ which are not leaves in $T'$. Choose $y_i$ such that $y_j$ is not an ancestor of $y_i$ for every $j\in\{1,\dots,l\}-\{i\}$. Let $T$ be the graph obtained from $T'$ by adding $x$ as an isolated vertex, adding the arc $(y_i,x)$, and for every $j\in\{1,\dots,l\}$, for every arc $(y_j,z)\in T$ with $z\in N^+(x)$, delete the arc $(y_j,z)$ and add the arc $(x,z)$. As $y_i$ is not reachable in $T'$ from any vertex $y\in N^-(x)-y_i$, there is no cycle in $T$. Hence $T$ is an outbranching of $D$ rooted at $r$ with at least $k$ leaves. Moreover, deciding the existence of a cut vertex and finding one if such exists can be done in polynomial time.
\item[(2)] Let $P$ be a bipath of length 4. Let $u$, $x$, $y$, $z$ and $t$ be the vertices of $P$ in this consecutive order. Let $D'$ be the rooted digraph obtained from $D$ by contracting $x$ and $y$. Let $T$ be an outbranching of $D$. Let $T'$ be the rooted digraph obtained from $T$ by contracting $y$ with its father in $T$. $T'$ is an outbranching of $D'$ with as many leaves as $T$. Let $T'$ be an outbranching of $D'$. If the father of $xy$ in $T'$ is $z$, then $T'-(z,xy)\cup (z,y) \cup (y,x)$ is an outbranching of $D$ with at least as many leaves as $T'$. If the father of $xy$ in $T'$ is $u$, then $T'-(u,xy)\cup (u,x) \cup (x,y)$ is an outbranching of $D$ with at least as many leaves as $T'$.
\item[(3)] Let $x$ be a vertex of $D$. Let $y\in N^-(x)$ be a vertex such that $N^-(x)-y$ cuts $y$ from $r$. Let $D'$ be the rooted digraph obtained from $T$ by deleting the arc $(y,x)$. Every outbranching of $D'$ is an outbranching of $D$. Let $T$ be an outbranching of $D$ containing $(y,x)$. There exists a vertex $z\in N^-(x)-y$ which is an ancestor of $x$. Thus $T-(y,x)\cup (z,x)$ is an outbranching of $D'$ with at least as many leaves as $T$.
\end{itemize}
$\square$


We apply these rules iteratively until reaching a \emph{reduced instance}, on which none can be applied.
\begin{lemma}\label{largedegree}
Let $D$ be a reduced rooted digraph with a vertex of indegree at least $k$. Then $D$ is a TRUE instance. 
\end{lemma}
\emph{Proof}: Assume $x$ is a vertex of $D$ with in-neighbourhood $N^-(x)=\{u_1,\dots,u_l\}$, with $l\ge k$. For every $i\in\{1,\dots,l\}$, $N^-(x)-u_i$ does not cut $u_i$ from $r$. Thus there exists a path $P_i$ from $r$ to $u_i$ outside $N^-(x)-u_i$. The rooted digraph $D'=\cup_{i\in\{1,\dots,l\}} P_i$ is connected, and for every $i\in\{1,\dots,l\}$, $u_i$ has outdegree 0 in $D'$. Thus $D'$ has an outbranching with at least $k$ leaves, and such an outbranching can be extended into an outbranching of $D$ with at least as many leaves. $\square$

\section{Quadratic kernel}\label{skernel}

In this section and the following, a vertex of a 2-connected rooted digraph $D$ is said to be \emph{special} if it has indegree at least 3 or if one of its incoming arcs is simple. A non special vertex is a vertex $u$ which has exactly two in-neighbours, which are also outneighbours of $u$. A \emph{weak bipath} is a maximal connected set of non special vertices. If $P=\{x_1,\dots,x_l\}$ is a weak bipath, then the in-neighbours of $x_i$, for $i=2,\dots,l-1$ in $D$ are exactly $x_{i-1}$ and $x_{i+1}$. Moreover, $x_1$ and $x_l$ are each outneighbour of a special vertex. Denote by $s(P)$ the in-neighbour of $x_1$ which is a special vertex.

This section is dedicated to the proof of the following statement:

\begin{theorem}\label{kernel}
A digraph $D$ of size at least $(3k-2)(30k-2)$ reduced under the reduction rules of previous section has an outbranching with at least $k$ leaves.
\end{theorem}

\emph{Proof}:
By Theorem~\ref{bound1} and Theorem~\ref{bound2}, if there are at least $6k+24k-1$ special vertices, then $D$ has an outbranching with at least $k$ leaves. Assume that there are at most $30k-2$ special vertices in $D$. 

As $D$ is reduced under Rule (2), there is no bipath of length 4. We can associate to every weak bipath $B$ of $D$ of length $t$ a set $A_B$ of $\lceil{t/3}\rceil$ out-arcs toward special vertices. Indeed,  let $P=(x_1,\dots, x_l)$ be a weak bipath of $D$. For every three consecutive vertices $x_i,x_{i+1},x_{i+2}$ of $P$, $2\le i \le l-3$, $(x_{i-1},x_i,x_{i+1},x_{i+2},x_{i+3})$ is not a bipath by Rule (2), hence there exists an arc $(x_j,z)$ with $j=i,i+1$ or $i+2$ and $z\notin P$. Moreover $z$ must be a special vertex as arcs between non-special vertices lie within their own weak bipath. The set of these arcs $(x_j,z)$ has the presribed size.

By Lemma~\ref{largedegree}, any vertex in $D$ has indegree at most $k-1$ as $D$ is reduced under Rule (3), hence there are at most $3(k-1)(30k-2)$ non special vertices in $D$.
$\square$

\vspace{12pt}

To sum up, the kernelization algorithm is as follows: starting from a rooted digraph $D$, apply the reduction rules. Let $D'$ be the obtained reduced rooted digraph. If $D$ has size more than $(3k-2)(30k-2)$, then reduce to a trivially TRUE instance. Otherwise, $D'$ is an instance equivalent to $D$ of size quadratic in $k$.

Our analysis for this quadratic kernel for \RMO is actually tight up to a constant factor.
Indeed, the following graph $T_l$ is reduced under the reduction rules stated on Section~\ref{srules} and has a number of vertices quadratic in its maximal number of leaves. Let $V=\{v_{i,j}|i=1,\dots,l,$ $j=1,\dots,3(l-1)\}$. For every $i=1,\dots,l$, $(r,v_{i,1})$ is an arc of $T$. For every $j=1,\dots,3l-2$, $i=1,\dots,l$, $(v_{i,j},v_{i,j+1})$ is a 2-circuit of $T_l$. For every $i=1,\dots,l$, $(v_{i,3l-1},v_{i+1 [l],3l-1})$ is an arc of $T_l$. For every $t=1,\dots,l-1$, $i=1,\dots,l$, $(v_{i,3t},v_{i+t[l],1})$ is an arc of $T_l$. This digraph $T_l$ is reduced under the reduction rules of Section~\ref{srules}, and $maxleaf(T_l)=2(l-1)$. Finally, $T_l$ has $3l(l-1)+1$ vertices.

Note that this graph has many 2-circuits. We are not able to deal with them with respect to kernelization. For the approximation on the contrary, we are able to deal with the 2-circuits to produce a constant factor approximation algorithm.

\section{Approximation}\label{approx}
Let us first point out that the reduction rules described in Section~\ref{rules} directly give an approximation algorithm asymptotically as good as the best known approximation algorithm \cite{DrescherMaxLeaf}. Indeed, as these rules are independant of the parameter, and as our proof of the existence of a solution of size $k$ when the reduced graph has size more than $3(k-1)(30k-2)$ is contructive, this yields a $O(\sqrt{OPT})$ approximation algorithm. Let us sketch this approximation algorithm. Start by applying the reduction rules described in Section~\ref{rules} to the input rooted digraph. This does not change the value of the problem. Let $m$ be the size of the reduced graph. Exhibit an outbranching with at least $\sqrt{\frac{m}{90}}$ leaves as in the proof of Theorem~\ref{kernel}. Finally, undo the sequence of contractions yield by the application of reduction rules at the start of the algorithm, repairing the tree as in the proof of Lemma~\ref{rules}. The tree thus obtained has at least $\sqrt{\frac{m}{90}}$ leaves, while the tree with maximum number of leaves in the input graph has at most $m-1$ leaves. Thus this algorithm is an $O(\sqrt{OPT})$ approximation algorithm.

Let us describe now our constant factor approximation algorithm for \RMOD being understood that this also gives an approximation algorithm of the same factor for \MO ~as well as for finding an out-tree (not necessarily spanning) with many leaves in a digraph.

Given a rooted digraph $D''$, apply exhaustively Rule (1) of Section~\ref{srules}. The resulting rooted digraph $D$ is 2-connected. By Lemma~\ref{rules}, $maxleaf(D'')=maxleaf(D)$.

Let us denote by $D_{ns}$ the digraph $D$ restricted to non special vertices. Recall that $D_{ns}$ is a dijoint union of bipaths, which we call \emph{non special components}. A vertex of outdegree 1 in $D_{ns}$ is called an \emph{end}. Each end has exactly one special vertex as an in-neighbour in $D$.

\begin{theorem}\label{majbound} 
Let $D$ be a 2-connected rooted digraph with $l$ special vertices and $h$ non special components. Then max$(\frac{l}{30},h-l)\le maxleaf(D)\le l+2h$.
\end{theorem}
\emph{Proof}: The upper bound is clear, as at most two vertices in a given non special component can be leaves of a given outbranching. The first term of the lower bound comes from Theorem~\ref{bound1} and Theorem~\ref{bound2}. To establish the second term, consider the digraph $D'$ which vertices are the special vertices of $D$ and $r$. For every non special component of $D$, add an edge in $D'$ between the special in-neighbours of its two ends. Consider an outbranching of $D'$ rooted at $r$. This outbranching uses $l-1$ edges in $D'$, and directly corresponds to an out-tree $T$ in $D$. Extend $T$ into an outbranching $\tilde{T}$ of $D$. Every non special component which is not used in $T$ contributes to at least a leaf in $\tilde{T}$, which concludes the proof.
$\square$

\vspace{12pt}

Consider the best of the three outbranchings of $D$ obtained in polynomial time by Theorem~\ref{bound1}, Theorem~\ref{bound2} and Theorem~\ref{majbound}. This outbranching has at least max$(\frac{l}{30},h-l)$ leaves. The worst case is when $\frac{l}{30}=h-l$. In this case, the upper bound becomes: $\frac{92l}{30}$, hence we have a factor $92$ approximation algorithm for \RMOB

\section{Conclusion}
We have given a quadratic kernel and a constant factor approximation algorithm for \RMOC reducing the gap between the problem of finding trees with many leaves in undirected and directed graphs. \ML ~ has a factor 2 approximation algorithm, and \RMO now has a factor 92 approximation algorithm. Reducing this 92 factor into a small constant is one challenge. The gap now essentially lies in the fact that \ML ~has a linear kernel while \RMO has a quadratic kernel. Deciding whether \RMO has a linear kernel is a challenging question. Whether long paths made of 2-circuits can be dealt with or not might be key to this respect.

\bibliographystyle{plain}
\bibliography{maxleafarxiv}

\begin{thebibliography}{10}

\bibitem{AlonMaxLeaf}
N.~Alon, F.~Fomin, G.~Gutin, M.~Krivelevich, and S.~Saurabh.
\newblock Parameterized algorithms for directed maximum leaf problems.
\newblock In {\em Proc. ICALP 2007, LNCS 4596}, pages 352--362, 2007.

\bibitem{AlonMaxLeaf2}
N.~Alon, F.~Fomin, G.~Gutin, M.~Krivelevich, and S.~Saurabh.
\newblock Spanning directed trees with many leaves.
\newblock {\em SIAM J. Discrete Maths.}, 23(1):466--476, 2009.

\bibitem{NoKernel}
H.~L. Bodlaender, R.~G. Downey, M.~R. Fellows, and D.~Hermelin.
\newblock On problems without polynomial kernels (extended abstract).
\newblock In {\em Proc. of Automata, Languages and Programming, 35th
  International Colloqium (ICALP)}, pages 563--574, 2008.

\bibitem{BonsmaMaxLeaf}
Paul~S. Bonsma and Frederic Dorn.
\newblock An fpt algorithm for directed spanning k-leaf.
\newblock abs/0711.4052, 2007.

\bibitem{stnumdir}
J.~Cheriyan and J.~Reif.
\newblock Directed s-t numberings, rubber bands, and testing digraph k-vertex
  connectivity.
\newblock {\em Combinatorica}, 14(4):435--451, 1994.

\bibitem{GutinMaxLeaf}
J.~Daligault, G.~Gutin, E.~J. Kim, and A~Yeo.
\newblock Fpt algorithms and kernels for the directed k-leaf problem.
\newblock {\em manuscript, http://arxiv.org/abs/0810.4946}, 2747, 2008.

\bibitem{MaxLeafApplication2}
E.~Dijkstra.
\newblock Self-stabilizing systems in spite of distributed control.
\newblock {\em Commun. ACM}, 17(11):643--644, 1974.

\bibitem{SeymourMaxLeaf}
G.~Ding, T.~Johnson, and P.~Seymour.
\newblock Spanning trees with many leaves.
\newblock {\em J. Graph Theory}, 37(4):189--197, 2001.

\bibitem{DowneyFellows}
R.G. Downey and M.R. Fellows.
\newblock {\em Parameterized complexity}.
\newblock Springer, 1999.

\bibitem{DrescherMaxLeaf}
M.~Drescher and A.~Vetta.
\newblock An approximation algorithm for the maximum leaf spanning arborescence
  problem.
\newblock {\em to appear in ACM Transactions on Algorithms}.

\bibitem{EstivillMaxLeaf}
V.~Estivill-Castro, M.~Fellows, M.~Langston, and F.~Rosamond.
\newblock Fixed-parameter tractability is polynomial-time extremal structure
  theory i: The case of max leaf.
\newblock In {\em Proc. of ACiD 2004}.

\bibitem{FernauMaxLeaf}
H.~Fernau, F.~Fomin, D.~Lokshtanov, D.~Raible, S.~Saurabh, and Y.~Villanger.
\newblock Kernel(s) for problems with no kernel: On out-trees with many leaves.
\newblock In {\em Proc. of STACS 2009}.

\bibitem{FlumGrohe}
J.~Flum and M.~Grohe.
\newblock {\em Parameterized Complexity Theory}.
\newblock Springer, 2006.

\bibitem{FominKratschMaxLeaf}
F.~Fomin, F.~Grandoni, and D.~Kratsch.
\newblock Solving connected dominating set faster than $2^n$.
\newblock {\em Algorithmica}, 52(2):153--166, 2008.

\bibitem{ApxHardMaxLeaf}
G.~Galbiati, F.~Maffioli, and A.~Morzenti.
\newblock A short note on the approximability of the maximum leaves spanning
  tree problem.
\newblock {\em Inf. Process. Lett.}, 52(1):45--49, 1994.

\bibitem{GuoNiedFPT}
J.~Guo and R.~Niedermeier.
\newblock Invitation to data reduction and problem kernelization.
\newblock {\em SIGACT News}, 38(1):31--45, 2007.

\bibitem{KneisMaxLeaf}
J.~Kneis, A.~Langer, and P.~Rossmanith.
\newblock A new algorithm for finding trees with many leaves.
\newblock In {\em Proc. of ISAAC 2008}.

\bibitem{stnum}
A.~Lempel, S.~Even, and I.~Cederbaum.
\newblock An algorithm for planarity testing of graphs.
\newblock In {\em In Theory of Graphs: Internat. Sympos.: Rome}, pages
  215--232. P. Rosenstiehl, Ed., 1966.

\bibitem{LLWemb}
N.~Linial, L.~Lovasz, and A.~Wigderson.
\newblock Rubber bands, convex embeddings and graph connectivity.
\newblock {\em Combinatorica}, 8:91--102, 1988.

\bibitem{NiedBookFPT}
R.~Niedermeier.
\newblock {\em Invitation to fixed parameter algorithms}, volume~31 of {\em
  Oxford Lectures Series in Mathematics and its Applications}.
\newblock Oxford University Press, 2006.

\bibitem{SolisMaxLeafApprox}
R.~Solis-Olba.
\newblock 2-approximation for finding trees with many leaves.
\newblock In {\em Proc. of ESA 1998}, pages 441--452.

\bibitem{StorerMaxLeaf}
J.~A. Storer.
\newblock Constructing full spanning trees for cubic graphs.
\newblock {\em Inform Process Lett}, 13:8--11, 1981.

\bibitem{MaxLeafApplication}
J.~Wu and H.~Li.
\newblock On calculating connected dominating set for efficient routing in ad
  hoc wireless networks.
\newblock In {\em DIALM '99: Proceedings of the 3rd international workshop on
  Discrete algorithms and methods for mobile computing and communications},
  pages 7--14, New York, NY, USA, 1999. ACM.

\end{thebibliography}

\end{document}